# Computational Fermi level engineering and doping-type conversion of $Ga_2O_3$ via three-step synthesis process

Anuj Goyal[1], Andriy Zakutayev[1], Vladan Stevanović[1,2], Stephan Lany[1]

[1]National Renewable Energy Laboratory, Golden, CO, 80401, United States.
[2]Colorado School of Mines, Golden, CO, 80401, United States

**Abstract**

$Ga_2O_3$ is being actively explored for high-power and high-temperature electronics, deep-ultraviolet optoelectronics, and other applications due to its ultra-wide bandgap and low projected fabrication cost of large-size and high-quality crystals. Efficient *n*-type doping of $Ga_2O_3$ has been achieved, but *p*-type doping faces fundamental obstacles due to compensation, deep acceptor levels, and the polaron transport mechanism of free holes. However, aside from achieving *p*-type conductivity, plenty of opportunity exists to engineer the position of the Fermi level for improved design of $Ga_2O_3$ based devices. We use first-principles defect theory and defect equilibrium calculations to simulate a 3-step growth-annealing-quench synthesis protocol for hydrogen assisted Mg doping in β-$Ga_2O_3$, taking into account the gas phase equilibrium between $H_2$, $O_2$ and $H_2O$, which determines the H chemical potential. We predict $Ga_2O_3$ doping-type conversion to a net *p*-type regime after growth under reducing conditions in the presence of $H_2$ followed by O-rich annealing, which is a similar process to the Mg acceptor activation by H removal in GaN. For equilibrium annealing with re-equilibration of compensating O vacancies, there is an optimal temperature that maximizes the $Ga_2O_3$ net acceptor density for a given Mg doping level, which is further increased in the non-equilibrium annealing scenario without re-equilibration. After quenching to operating temperature, the $Ga_2O_3$ Fermi level drops below mid-gap down to about +1.5 eV above the valence band maximum, creating a significant number of uncompensated neutral $Mg_{Ga}^0$ acceptors. The resulting free hole concentration in $Ga_2O_3$ is very low even at elevated operating temperature (~$10^8$ cm$^{-3}$ at 400C) due to deep energy level of these Mg acceptors, and hole conductivity is further impeded by the polaron hopping mechanism. However, the Fermi level reduction down to +1.5 eV and suppression of free electron density in this doping type converted ($N_A > N_D$) $Ga_2O_3$ material is of significance and impact for the design of $Ga_2O_3$ power electronics devices.

Monoclinic beta-phase of gallium oxide (β-Ga$_2$O$_3$) an exciting material for various (opto)electronic and energy-related technologies due to its ultra-wide bandgap (~4.9 eV) and the ability to grow large-size high-quality single crystals at low projected cost.[1,2]. Considered applications include various power electronic devices[3], radio-frequency transistors[4,5], solar-blind photo-detectors[6–8], gas sensors[9,10], contact layers in photovoltaics[11,12], and other applications[13,14]. In power electronics, dedicated efforts in recent years towards improving crystal growth, *n*-type doping, and device fabrication led to substantial progress in performance of β-Ga$_2$O$_3$ based horizontal transistors, vertical Schottky barrier diodes (SBDs), vertical metal oxide semiconductor field effect transistors (MOSFETs), and related devices[15–17]. For example, depletion mode (normally-on) current aperture vertical β-Ga$_2$O$_3$ MOSFETs with implantation doping[17] and enhancement-mode (normally-off) Ga$_2$O$_3$ vertical transistors with fin-shaped channels[18] have been demonstrated. However, one of the biggest challenges in realizing the true potential of Ga$_2$O$_3$ in power electronics, along with its low thermal conductivity, is the absence of *p*-type doping, limiting the design of device structures that can be realized[16].

Ga$_2$O$_3$ is intrinsically an *n*-type semiconductor. Using extrinsic donors its *n*-type conductivity is easily tunable over many orders of magnitude[1,19–23], but *p*-type doping faces fundamental obstacles[24–31]. Among various acceptor dopants in Ga$_2$O$_3$, Mg is computationally predicted to be the most stable[27] and was also experimentally found to reduce the unintentional *n*-type conductivity and increase the resistivity of the material[26]. Mg doped Ga$_2$O$_3$ has been synthesized[32] and studied[33] using an electron paramagnetic resonance (EPR) technique, where Mg acceptor (0/1-) level is experimentally determined to be at $E_V$+0.7 eV. However, in a photoluminescence study[34], the Mg acceptor level is shown to be deeper at $E_V$+1.0 eV, which compares better with the theoretical predictions employing hybrid functional calculations[25,28,35,36], where (0/1-) level is estimated to be between $E_V$+1.0 to 1.5 eV. The differences in theoretical predictions published in literature originate to some extent from different fraction of exact (non-local) exchange employed in these hybrid functional calculations. In more recent computational work, focus has been towards studying diffusion[28], EPR[37] and luminescence properties[38] of Mg doped Ga$_2$O$_3$.

While the prospects of achieving *p*-type conductivity through acceptor doping in β-Ga$_2$O$_3$ remain bleak, ample opportunities exist for Fermi level engineering of Ga$_2$O$_3$ power electronic devices via acceptor-type dopants[1,16,26]. Even when holes remain localized at the acceptor site or as small polarons, the doping can cause the acceptor concentration ($N_A$) to exceed the donor concentration ($N_D$). The result is a doping type conversion with a large drop of the Fermi level $E_F$ where electrons become minority carriers. The resulting acceptor-doped Ga$_2$O$_3$ material can be used as buried electron barrier, a.k.a. current blocking layer, for controlling the turn-on voltage and saturation current of vertical metal-oxide-semiconductor field effect transistors (MOSFET) with current aperture. Such acceptor-doped Ga$_2$O$_3$ can be also used for increasing the breakdown voltage and decreasing the leakage current of vertical MOSFET and Schottky diodes using edge termination by guard rings and related structures[1,16,17,26,39]. It is likely that such acceptor doped β-Ga$_2$O$_3$ can be achieved using non-equilibrium synthesis process with mobile hydrogen, species that was instrumental in achieving *p*-type doping of GaN, which led to the development of the blue light emitting diode.[40–44] More recently, non-equilibrium processing involving hydrogen and oxygen was employed to reduce the electron density in otherwise degenerately doped ZnSnN$_2$[45,46] and MgZrN$_2$.[47]

Here we study H-assisted Mg doping of β-Ga$_2$O$_3$ under non-equilibrium growth and annealing process, in a broad analogy to the well-known strategies for activating *p*-type conductivity in Mg-doped GaN thin films, using first principles supercell calculations and thermodynamic defect equilibrium simulations. Although traditional *p*-type semiconductivity resulting from thermal ionization of free holes at room temperature in Ga$_2$O$_3$ remains beyond reach, we report quantitative computational predictions for synthesis process conditions that enable doping type conversion in Ga$_2$O$_3$. We predict doping-type conversion to a net *p*-type regime after O-rich annealing of Ga$_2$O$_3$ grown under reducing conditions in the presence of H$_2$. The resulting doping type-converted Ga$_2$O$_3$ has Fermi level +1.5 eV above the valence band maximum due to uncompensated neutral Mg$_{Ga}^0$ acceptors, but the resulting free hole concentration is very low (~10$^8$ cm$^{-3}$ even at 400C) due to deep energy level of these acceptors. These theoretical predictions are expected to guide experimental tuning of growth and annealing conditions to achieve doping type converted Ga$_2$O$_3$ for realization of novel device configurations.

As illustrated in Figure 1, our approach consists of a three-step process of (1) thin-film growth under O-poor conditions in the presence of hydrogen, (2) acceptor activation via annealing in an O-rich/H-poor atmosphere, and (3) quenching to a range of operating temperatures. The thermodynamic simulations take into account the H$_2$ + ½O$_2$ ↔ H$_2$O gas phase equilibrium connecting the O and H chemical potentials. In the growth step, the presence of H donor impurities can increase the solubility of substitutional Mg$_{Ga}$ acceptors, and it reduces the concentrations of compensating O vacancies ($V_O$) while the system remains *n*-type. In the annealing step, meant to purge the mobile H donor species, we consider two different stages of non-equilibrium. In the more readily realizable scenario, only the equilibration of the Mg solubility is suppressed. In the second scenario, which might be more difficult to realize, also the equilibration of $V_O$ formation is suppressed, thereby allowing for maximal non-equilibrium activation of acceptor dopants and ensuing Fermi level reduction. In the quenching step, it is assumed that all dopant and defect concentrations remain at the level of the preceding process step, and only the Fermi level ($E_F$) and the corresponding electron and hole densities are equilibrated to the operating temperature of interest.

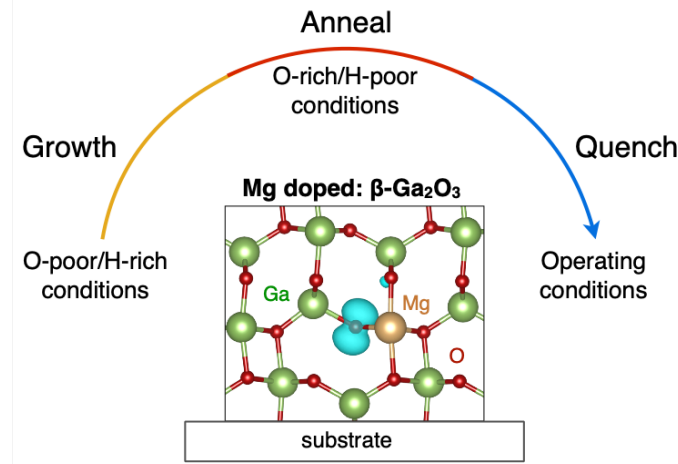

Figure 1. Schematic illustration of thermodynamic conditions in the modeling of growth, annealing, and quenching steps of Mg-doped β-Ga$_2$O$_3$. The atomic structure model shows

the spin-density iso-surface (blue) associated with the defect-bound hole-polaron created by the neutral $Mg_{Ga}^0$ acceptor at a neighboring O site.

*Defect calculations*. The stable phase, $\beta$-$Ga_2O_3$, has a monoclinic crystal structure (space group C2/m) with two non-equivalent crystallographic Ga sites (Ga1 tetrahedral, and Ga2 octahedrally coordinated) and three O sites (O1 bonded to one Ga1 and two Ga2, O2 bonded to two Ga1 and one Ga2, and O3 bonded to one Ga1 and three Ga2). Figure 1 shows the hole charge density of the $Mg_{Ga}$ defect (from hybrid functional calculations) in the neutral charge state in the $Ga_2O_3$ crystal structure illustrating the "polaronic" deep nature of $Mg_{Ga}$ acceptor defect level. Mg prefers the six-fold coordinated Ga2 site, and the acceptor-hole is localized on the neighboring O1 site. Our simulated description of $Mg_{Ga}$ acceptor is consistent with the EPR study[48] that also found the hole to be located at the O1 nearest neighbor oxygen site.

We calculated the acceptor levels for the two $Mg_{Ga}$ configurations using a hybrid functional (see below for computational details). The formation energies $\Delta H_D$ of dopants, defects, dopant-defect pairs and complexes were calculated using the approach of Ref[49], combining density functional theory (DFT) supercell energies with *GW* band gap corrections and fitted elemental reference energies[50] (FERE). We included the native defects ($V_{Ga}$, $V_O$), extrinsic dopants ($Mg_{Ga}$, $Mg_i$, $H_i$) and defect pairs and complexes ($Mg_{Ga}+V_O$, $2Mg_{Ga}+V_O$, $Mg_{Ga}+H_i$) expected to form due to charge compensation. H can also substitute for oxygen ($H_O$),[22] but we find that the formation energy of $H_O$ is 0.5 eV higher than that of $H_i$ even under the extreme limit of O poor conditions ($\Delta\mu_{Ga} = 0$, $\Delta\mu_O = -3.75$ eV), and we will therefore not further consider $H_O$ in this study.

Table 1 lists the defect formation energies and the binding energies $\Delta E_b$ of defect pairs, where the formation energies are given for the convention $E_F = E_{VBM}$ and $\Delta\mu_i = 0$, and the actual values of $\Delta H_D$ at any given point in the thermodynamic simulation (see below) are determined for the actual Fermi level position and chemical potentials.[49] The formation energy of the $Mg_{Ga}^0$ defect on the Ga2 site is lower by 0.43 eV compared to Mg substituting the Ga1 site, and the (0/1−) charge transition level is calculated at 1.0 and 1.1 eV above the valence band maximum (VBM) for Mg on the Ga2 and Ga1 site, respectively. Oxygen vacancies in $Ga_2O_3$ act as deep donors, with a (2+/0) charge transition ~1.3 eV below the conduction band minimum (CBM), whereas the hydrogen interstitial act as shallow donor, stable in the 1+ charge state throughout the range of Fermi energies in the bandgap, consistent with prior literature results.[22,24] We find that the binding of the H interstitial to Mg acceptors is weaker than the binding of O vacancies (*cf*. Table 1), suggesting that H interstitials are more easily separated from the acceptors during annealing and activation than the O vacancies.

Table 1: Calculated formation energies $\Delta H_{ref}$ of defects and dopants and the binding energies $\Delta E_b$ of dopant-defect complexes. The formation energies are given for a reference condition (see Ref. 47), with the chemical potentials set to elemental reference ($\Delta\mu_i = 0$) and the Fermi energy set to the VBM.

| Defect | $\Delta H_{ref}$ (eV) | Defect complex | $\Delta E_b$ (eV) |
|---|---|---|---|
| $V_{O1}^{2+}$ | −2.53 | $(Mg_{Ga2} + V_{O1})^+$ | −0.74 |
| $V_{O2}^{2+}$ | −1.83 | $(Mg_{Ga2} + V_{O2})^+$ | −0.58 |
| $V_{O3}^{2+}$ | −2.66 | $(Mg_{Ga2} + V_{O3})^+$ | −0.62 |
| $V_{Ga1}^{3-}$ | +16.71 | $(Mg_{Ga2} + H_i)^0$ | −0.44 |
| $Mg_{Ga2}^{1-}$ | +2.12 | $(2Mg_{Ga2} + V_{O1})^0$ | −1.03 |
| $H_i^+$ | −3.60 | $(2Mg_{Ga2} + V_{O3})^0$ | −1.06 |

The calculated defect formation energies along with the binding energies of defect pairs are used as an input into our thermodynamic modeling[51,52], which yields quantitative defect concentrations, doping limits, and the position of the Fermi level $E_F$ as function of dopant (Mg) concentration, the partial pressures of oxygen ($pO_2$), hydrogen ($pH_2$) or water vapor ($pH_2O$), and the temperature ($T$). The Fermi level is obtained as a self-consistent solution that observes the charge balance between defect charges and free carriers, while the defect concentrations are obtained from the defect formation energies for the same $E_F$. The chemical potentials $\mu_i = \mu_i^0 + \Delta\mu_i$ are expressed by the deviation $\Delta\mu_i$ from to the elemental reference energy $\mu_i^0$. Using the FERE values of Ref.[50], we obtain the formation enthalpy of $Ga_2O_3$ as $\Delta H_f(Ga_2O_3) = 2\Delta\mu_{Ga} + 3\Delta\mu_O$ = −11.26 eV, which defines the relationship between Ga and O chemical potentials under the phase-coexistence of $Ga_2O_3$. Similarly, using the tabulated[53] formation enthalpy of water vapor, we have $\Delta H_f(H_2O) = -2.48$ eV $= 2\Delta\mu_H + \Delta\mu_O - \Delta\mu_{H2O}$, where the $\Delta\mu$ values for the gas phases are calculated as function of $T$ and partial pressure via the ideal gas equations with the tabulated standard enthalpies and entropies for 298K and 1 atm (note, $\Delta\mu_O$ = ½ $\Delta\mu_{O2}$ and $\Delta\mu_H$ = ½ $\Delta\mu_{H2}$). Similar thermodynamic simulations have been successfully employed in the past in other oxides to quantitatively predict carrier concentrations as a function of synthesis and measurement conditions, for example in case of the non-equilibrium origin of conductivity in Ga doped ZnO[54].

*Growth step:* In $Ga_2O_3$, as in other wide gap oxides, it is generally observed that annealing in oxygen rich environment reduces the free electron density, while annealing in hydrogen rich environment increases it[1]. This behavior reflects general doping principles[55], where O-rich conditions reduce the formation energy of electron compensating defects like cation vacancies and O interstitials. H-rich environments promote the formation of donor-like H interstitials[56] but at the same time, exposure to $H_2$ gas also creates O-poor reducing conditions due to the gas phase equilibrium with water vapor, a well-established process that is frequently used in solid state chemistry[57]. These reducing conditions generally favor *n*-type conductivity. According to the doping principles, the increased Fermi energy in *n*-type material reduces the formation energy of acceptor-type dopants, thereby enhancing their solubility. Finally, the mobile nature of H interstitials allows to exploit non-equilibrium processing, such as the removal of hydrogen by annealing while the equilibration of other processes remains suppressed, like the exsolution of acceptor dopants or the formation of compensating intrinsic defects[44]. These concepts are exploited in our initial growth step illustrated in Fig. 2.

Figure 2a shows the equilibrium solubility of Mg as function of temperature and $pO_2$. Under equilibrium conditions, Hydrogen addition is equivalent to maintaining a certain partial pressure $pH_2O$ of water vapor, and the hydrogen chemical potential ($\Delta\mu_H$) is determined by the gas phase equilibrium as described above. In Fig. 2a, we used $pH_2O = 10^{-5}$ atm as an estimated upper bound for a typical physical vapor deposition (PVD) process such as molecular beam epitaxy (MBE), but the solubility of Mg is barely affected at this low level of hydrogen addition. The Mg equilibrium solubility increases with increasing temperature and decreasing $pO_2$ (both of which reduce $\Delta\mu_O$) as expected from the doping principles and remains in the sub-percent range except under extreme reducing conditions.

Figure 2b shows defect equilibria as function of the Mg doping level and the hydrogen chemical potential $\Delta\mu_H$, which is controlled by $pH_2O$ (*cf.* left and right vertical axes in Fig 2b), assuming growth conditions at $T_g$ = 600 °C and $pO_2$ = $10^{-9}$ atm, as representative for low $T$ and moderate $pO_2$ conditions for PVD (MBE) growth of $Ga_2O_3$[58]. The solid black line in Fig. 2b marks the equilibrium thermodynamic solubility limit of Mg in $Ga_2O_3$ determined by the competing phase $MgGa_2O_4$. Within the range of data shown in Fig 2b, the Mg chemical potential does not exceed the bounds determined by $Mg(OH)_2$, $MgH_2$, MgO, and metallic Mg. We observe that significant increases of the Mg solubility occur at $pH_2O$ > $10^{-5}$ atm, reaching about Mg/(Mg+Ga) = 0.1% at $pH_2O$ = 1 atm, which is feasible in atmospheric pressure chemical vapor deposition processes (APCVD) of β-$Ga_2O_3$[59,60]. Even higher Mg solubilities could result from using an activated source, where the H chemical potential $\Delta\mu_H$ can exceed the thermodynamic limits. Such hydrogen plasma has been successfully used in the past during atomic layer deposition (ALD) growth of AlN[61] and for SiC surface cleaning prior to molecular beam epitaxy (MBE) growth of GaN[62]. In Figure 2b and the following discussion we consider non-equilibrium Mg concentrations up to 10%, since in thin-film growth the thermodynamic solubility limit can often be exceeded due to slow exsolution kinetics[63]. For experimental reference, we note that in an ion implantation study[26] Mg dopant concentration up to $1.5\times10^{19}$ $cm^{-3}$ (about 0.05% of the $3.82\times10^{22}$ $cm^{-3}$ Ga sites) has been achieved in bulk substrates, and it is expected that still much higher concentrations are achievable in thin film growth.

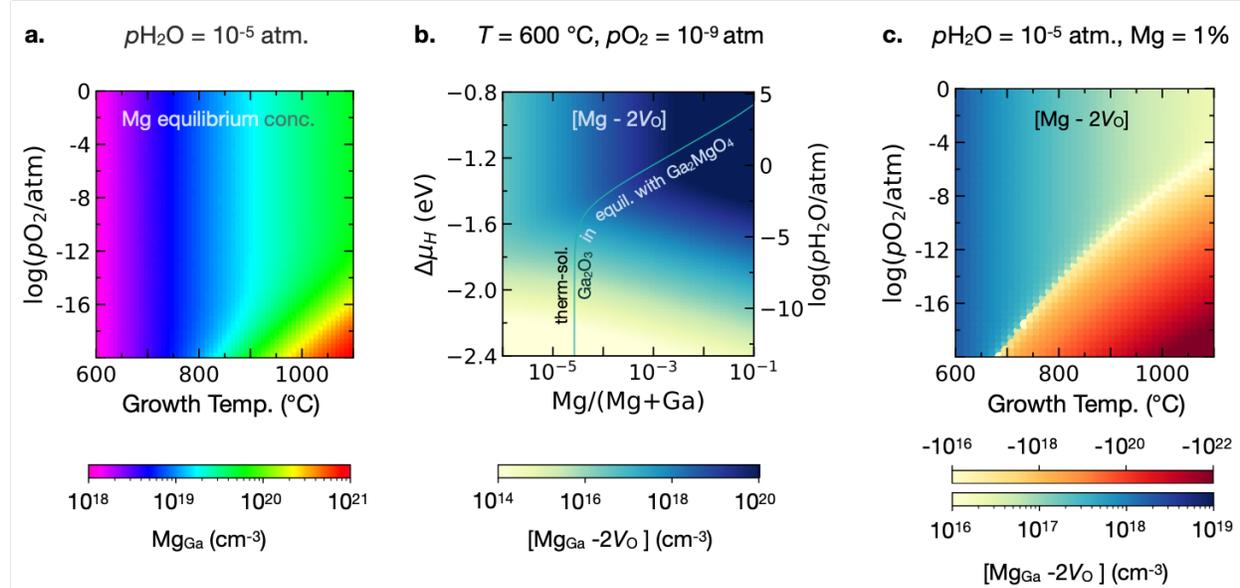

Figure 2. *Growth step*: (a) Predicted $Mg_{Ga}$ defect concentration as function of growth temperature ($T_g$) and $O_2$ partial pressure ($pO_2$). (b) Defect equilibria as a function of the Mg/(Mg+Ga) ratio and the water vapor partial pressure ($pH_2O$) under O-poor growth conditions typical for thin film growth. The solid line shows the Mg solubility limit and the color scale shows the dopant-defect difference concentration [$Mg_{Ga}$ − 2$V_O$]. (c) [$Mg_{Ga}$ − 2$V_O$] as function of $T_g$ and $pO_2$ at 1% Mg doping.

Besides Mg solubility considerations, growth under hydrogen addition could also be utilized to minimize the contribution of intrinsic defects to the acceptor compensation[43]. If the compensation is dominated by $H_i$ donors, the Mg acceptors could be activated by annealing out the hydrogen, proposed that the intrinsic defects would not re-equilibrate (see discussion below). To quantify this effect, we define the dopant-defect difference concentration [$Mg_{Ga} - 2V_O$] with the results shown as a color scale in Figure 2b. The significance of this number is that it equals the hypothetical net acceptor concentration if all hydrogen could be purged during the high temperature annealing step without introducing additional $V_O$ defects. Therefore, we will refer to it as the "precursor" acceptor concentration. The color scale in Fig 2b shows [$Mg_{Ga} - 2V_O$] increasing moderately with Mg composition and rather strongly with $pH_2O$, and the associated H chemical potential. Thus, the hydrogen serves to reduce the intrinsic compensation mechanism, thereby maximizing the precursor acceptor concentration for the subsequent annealing step. Figure 2c explores the dependence of the [$Mg_{Ga} - 2V_O$] concentration as function of $T$ and $pO_2$ at fixed $pH_2O$ and Mg composition. We observe that this precursor acceptor density decreases with increasing growth temperature. Also, for too reducing conditions (too high $T_g$ or too low $pO_2$), [$Mg_{Ga} - 2V_O$] becomes negative, implying that the system would remain $n$-type even if all H were removed. Thus, low temperature conditions are favorable for the initial growth step.

*Annealing step:* In the post-growth annealing step (Fig. 3), H-poor/O-rich conditions are considered, analogous to the H-poor/N-rich conditions used for Mg acceptor activation in GaN[40,41]. At 600°C, the O chemical potential increases from $\Delta\mu_O = -1.72$ to $-0.94$ eV as $pO_2$ increases from $10^{-9}$ atm in the growth step to 1 atm considered here for the O-rich annealing condition. Apart from increasing the formation energy of compensating O vacancies, the result is also a reduction of the H chemical potential from $\Delta\mu_H = -1.68$ (growth step with $pO_2 = 10^{-9}$ atm and $pH_2O = 10^{-5}$ atm) to $-2.33$ eV when taking $pH_2O = 10^{-8}$ atm (10 ppb) for purified $O_2$ at 1 atm used in the annealing step. Additionally, increasing the temperature during annealing lowers both $\Delta\mu_O$ and $\Delta\mu_H$, e.g., by about 0.4 eV between 600 and 900°C. Thus, the Mg acceptor activation relies on both suppressing compensation by $V_O$ and reduction of the H concentration during annealing. Kinetically, the annealing conditions must also be such that H is sufficiently mobile to diffuse out while Mg remains immobile so to prevent Mg segregation. The rationale for choosing the present process parameters comes from recent Mg doping studies[26,28,64], indicating that Mg needs $T >$ 800 °C to become mobile, and calculations of a small migration barrier of 0.34 eV[22] for $H_i^+$ indicating sufficiently fast H diffusion. A more subtle question is whether the annealing process can be performed such that the O vacancy concentrations from the growth step are maintained or whether they equilibrate. Since it could be feasible to realize either situation, we will consider both scenarios.

*Equilibrium annealing with $V_O$ equilibration*. In this scenario, only the total Mg concentration is carried over from the growth step, and all other degrees of freedom are equilibrated, while switching to the O-rich/H-poor regime ($pO_2 = 1$ atm, $pH_2O = 10^{-8}$ atm). Thus, in this scenario, the H addition of the growth step is relevant only in so far it supports the incorporation of Mg in the $Ga_2O_3$ lattice. Figure 3a shows the resulting net acceptor concentration [$Mg-2V_O-H_i$] obtained from thermodynamic modeling as function of annealing temperature. As the annealing temperature increases, H is purged out of the system, resulting in an increasing concentration of uncompensated net Mg acceptors, making the system net $p$-type in the sense of a positive value of the [Mg – 2$V_O$ – $H_i$] concentration difference. However, as the annealing temperature is increased, the O vacancy concentration also increases (driven by a reduction of $\Delta\mu_O$) and therefore, there is a tipping point with respect to the annealing temperature above which O vacancies compensate the Mg acceptors and make the system net *n*-type.

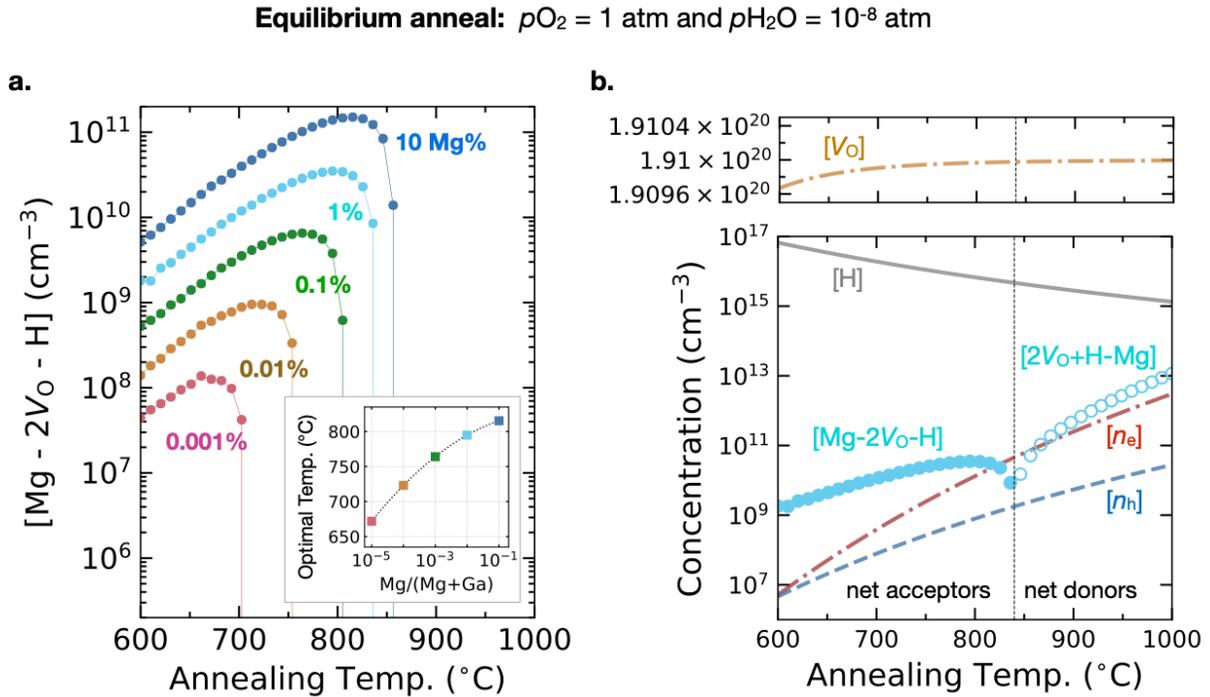

Figure 3. *Equilibrium annealing step with $V_O$ equilibration*: (a) Predicted net acceptor concentration [Mg – 2$V_O$ – $H_i$] for the O-rich/H-poor annealing step. The inset shows the dependence of the optimal annealing temperature $T_a$ on the Mg concentration. (b) The detailed defect equilibrium for Mg/(Mg+Ga) = 1%. The vertical dotted line indicates the crossover from net *p*-type to net *n*-type.

While the Fermi energy remains in the mid-gap region during annealing, it exhibits a small shift towards higher energies with increasing temperature, enough to affect the ratio between charged and ionized defects ($Mg_{Ga}^0$ vs $Mg_{Ga}^-$ and $V_O^0$ vs $V_O^{2+}$). These changes in the order of $10^{10}$ cm$^{-3}$ contribute to the charge balance and, ultimately, cause the crossover from net *p*-type to net *n*-type. As seen in the inset of Fig 3a, the optimum annealing temperature, i.e., the temperature at which the net *p*-type doping is maximized, increases from 670 to 820 °C with increasing Mg composition from 0.001% to 10%. The resulting net acceptor concentrations in the range of $10^8$ - $10^{11}$ cm$^{-3}$ are rather low, but the type conversion to net *p*-type is a significant hallmark, since it affords very low free electron densities at device operating conditions (see below). The resulting net acceptor density [Mg – 2$V_O$ – $H_i$] is much lower than the precursor acceptor density [Mg – 2$V_O$] determined above for the growth step. This incomplete acceptor activation is due to the fact that not all hydrogen is purged, but also due to an increase of the concentration of compensating O vacancies during the annealing step, resulting from a lowered formation energy of $V_O^{2+}$ as $E_F$ is reduced, even as $\Delta\mu_O$ is increased compared to the growth step.

(Hence, a higher net *p*-type doping is expected for the case when $V_O$ equilibration is suppressed, as discussed below.) The residual H concentration ranges from $4.3 \times 10^{16}$ to $4.5 \times 10^{17}$ cm$^{-3}$ at the optimal temperature for the Mg doping level between 0.001% and 10%, respectively.

*Non-equilibrium annealing without $V_O$ equilibration*. We are also considering possible non-equilibrium annealing conditions in $Ga_2O_3$, even though these might be more difficult to realize experimentally. Under the assumption that O diffusion is much slower than that of H, the annealing process could be performed such that H diffuses out while the O vacancy formation does not have enough time to re-equilibrate. In this scenario, the $V_O$ concentration is fixed to that of the preceding growth step, and the H addition during growth has higher significance than in the equilibrium annealing case, since it serves to reduce the compensation by $V_O$ and thereby increase the precursor acceptor density [Mg − 2$V_O$] (*cf*. Fig. 2b). The calculated values of O vacancy migration barriers in the literature[65] indeed indicate a much lower mobility compared to interstitial hydrogen[22]. Therefore, non-equilibrium annealing could be feasible if the annealing process is performed sufficiently fast. Even though the total $V_O$ concentration remains constant, the dopant-defect pair association-dissociation process is equilibrated in the simulation, because this process does not require long-range diffusion and should be comparatively fast.

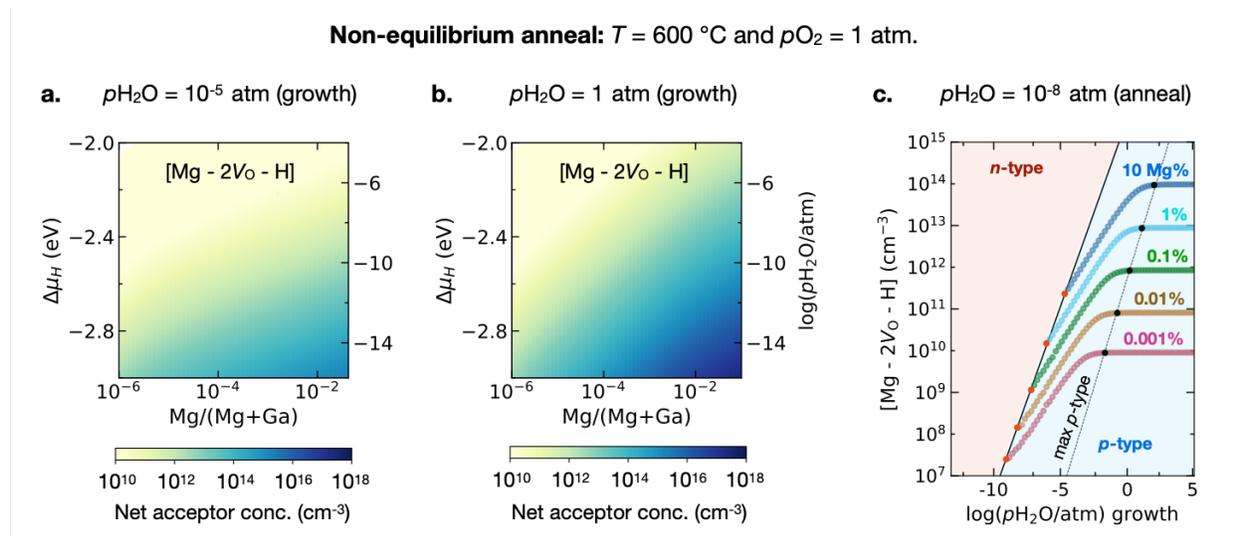

Figure 4. *Non-equilibrium annealing step without $V_O$ equilibration*: Predicted net acceptor concentration [Mg − 2$V_O$ − $H_i$] under O-rich/H-poor annealing conditions at $T_a$ = 600 °C and $pO_2$ = 1 atm, as function of Mg-doping level and $pH_2O$ (controlling $\Delta\mu_H$) during the anneal. Non-equilibrium conditions are assumed such that the O-vacancy concentration of the preceding growth step is maintained. The underlying growth step conditions are consistent with Fig. 2b, i.e., $T_g$ = 600 °C, $pO_2$ = 10$^{-9}$ atm, with $pH_2O$ = 10$^{-5}$ atm in (a) and $pH_2O$ = 1 atm in (b). The dependence of the net acceptor concentration on $pH_2O$ during the growth step is shown in (c).

Figure 4 shows the predicted net acceptor concentration [Mg − 2$V_O$ − $H_i$] for annealing without $V_O$ equilibration. The underlying growth step conditions are identical to those considered above (*cf*. Fig. 2b), i.e., $T_g$ = 600 °C, $pO_2$ = 10$^{-9}$ atm, with hydrogen exposure during growth taken as $pH_2O$ = 10$^{-5}$ atm for Fig. 4a and 1 atm for Fig 4b. For the annealing conditions, we maintain the same temperature, $T_a$ = 600 °C and use O-rich conditions with $pO_2$ = 1 atm as in the equilibrium annealing case above. According to migration barrier calculations,[65] O vacancies are beginning to become mobile at this temperature. Note that the model here assumes equilibration of O vacancies during growth but not during annealing, despite the fact that both processes occur at same temperature. This assumption can be justified by the general observations that the surface kinetics during growth is usually faster than in the bulk, and the diffusion distance required for equilibration is much larger for the completed film during annealing than it is during deposition[66,67].

We observe in Fig. 4 the expected increase of the net acceptor density with increasing Mg-doping level and decreasing H-chemical potential $\Delta\mu_H$, which is determined by the corresponding $pH_2O$ during the annealing. Compared to the annealing with $V_O$ equilibration, the net acceptor density increases considerably under otherwise identical conditions. For example, for growth with 1% Mg doping and $pH_2O = 10^{-5}$ atm, we obtain $[Mg - 2V_O - H_i] \approx 10^{11}$ cm$^{-3}$ (Fig 4a) after non-equilibrium annealing at 600 °C and $pH_2O = 10^{-8}$ atm (purified $O_2$), about 1-2 orders of magnitude higher than in case of annealing with $V_O$ equilibration (cf. Fig. 3a). Higher net acceptor densities, up to $10^{14}$ cm$^{-3}$ (Fig 4b), result when the preceding growth step is performed under very H rich conditions with $pH_2O = 1$ atm. These results suggest that H addition during growth is most impactful when choosing deposition techniques that allow for high $H_2O$ partial pressures. As seen in Fig 4c, a minimum $pH_2O$ must be supplied during growth to enable p-type conversion during non-equilibrium annealing, for example $pH_2O = 10^{-7}$ atm for 1% Mg doping. In the absence of $H_2O$ addition, the compensation is dominated by O vacancies during growth, preventing type conversion during annealing. The $H_2O$ partial pressure required to achieve maximal net p-type doping increases from $10^{-2}$ atm to $10^2$ atm with increasing Mg concentration from 0.001% to 10%. These results provide theoretical guidance for optimization of the post-growth annealing step of Mg-doped $Ga_2O_3$.

*Quenching step*: Finally, we are considering the doping situation under device operating conditions. It is assumed that the defect concentrations from the preceding annealing step are quenched-in, and no equilibration with the environment takes place (e.g., H uptake from ambient precluded by encapsulation). During this step, only the electronic degrees of freedom equilibrate, i.e., the Fermi level $E_F$ and the corresponding electron ($n_e$) and hole ($n_h$) concentrations. We are considering elevated device operating temperatures $T_o$, as relevant for $Ga_2O_3$ high-power and high-temperature electronics applications[68]. Figure 5a shows the Fermi level as function of $T_o$ and the Mg doping level for the case of $V_O$ equilibration during the preceding annealing step at the optimum temperature $T_a$ for any given Mg doping level (cf. Fig. 3 insert). Figures 5b and 5c are for the case of non-equilibrium annealing at 600 °C, where the growth step was performed at the same temperature with $pH_2O = 10^{-5}$ atm and 1 atm, respectively (cf. Fig. 4a and 4b). In all cases, the annealing conditions before the quench were taken at $pO_2 = 1$ atm and $pH_2O = 10^{-8}$ atm as discussed above.

In Fig. 5a, we observe a moderate reduction of $E_F$ with the Mg doping level. The Fermi level increases with operating temperature $T_o$, but remains within the lower half of the band gap (net p-type). While the absolute hole carrier concentration remains very low (below $10^6$ cm$^{-3}$), the more important finding is here that the electron concentration remains negligible (below $10^3$ cm$^{-3}$) up to 400 °C. The suppression of the electron density is an important feature of the type conversion, and our simulations suggest that the proposed growth-annealing process is suitable to effectively suppress the electron conductivity, e.g., for the purpose to minimize the associated leakage currents in device applications. However, achieving actual p-type conductivity is a much greater challenge. The quenching from the non-equilibrium annealing step, shown in Figs. 5b and 5c, affords a stronger Fermi level reduction than the annealing with $V_O$ equilibration. Even then, however, the corresponding hole concentrations are only about $10^8$ cm$^{-3}$ at $T_o$ = 400 °C (Fig. 5c). Since holes tend to form localized polarons rather than acting as band-like free carriers, the p-type conductivity is further limited by the hole mobility. Based on a predicted polaron hopping barrier of 0.4 eV, the mobility was estimated in Ref.[69] as $10^{-6}$ cm$^2$ V$^{-1}$ s$^{-1}$ at room temperature, which could increase to about $10^{-3}$ cm$^2$ V$^{-1}$ s$^{-1}$ at 400 °C, resulting in a conductivity σ ≈ $10^{-14}$ S/cm for the above mentioned hole density. These rough estimates illustrate the challenges for achieving significant p-type conductivity even at elevated temperatures and under optimistic assumptions about suppressing the $V_O$ compensation mechanism.

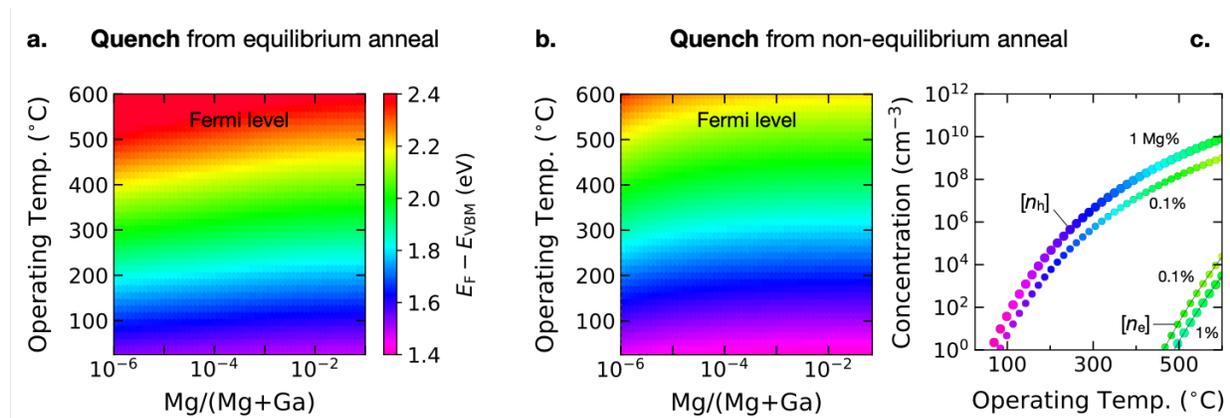

Figure 5. *Quench step*: (a) Predicted Fermi level ($E_F$ – $E_{VBM}$) as function of operating temperature $T_o$ and Mg doping level, after quenching-in the defect concentrations from the preceding annealing step with $V_O$ equilibration at the optimal $T_a$ for each Mg concentration (inset in Fig. 3a). (b) Fermi level after quench from the preceding non-equilibrium annealing step at T = 600°C with growth at $pH_2O$ = $10^{-5}$ atm (*cf*. Fig. 4a), and (c) with growth at $pH_2O$ = 1 atm (*cf*. Fig. 4b). In (c), the $T_o$ dependence of the electron [$n_e$] and hole [$n_h$] concentrations corresponding to the respective $E_F$ is shown for Mg concentrations of 0.1% and 1%.

*Conclusions*: We performed a thermodynamic simulation of a synthesis process (growth-annealing-quench sequence) for hydrogen-assisted magnesium acceptor doping in $Ga_2O_3$ based on defect energies obtained from first principles calculations. During the growth step, the thermodynamic solubility limit of Mg acceptor dopants is significantly increased for partial pressures above $pH_2O$ > $10^{-5}$ atm. The H exposure during growth also serves to reduce the acceptor compensation through $V_O$ defects. The annealing step at O-rich/H-poor conditions ($pO_2$ = 1 atm and $pH_2O$ = $10^{-8}$ atm) can produce a doping type conversion from net n- to net p-type, and we consider both cases with and without $V_O$ re-equilibration. For the case with $V_O$ equilibration, there is an optimal annealing temperature at which the net p-type doing is maximized, e.g., about 800 °C for 1% Mg doping, but too high annealing temperatures result in reversion to net n-type. The net p-type doping can be further increased if the $V_O$ equilibration can be suppressed, e.g., in rapid thermal processing at lower temperature (600 °C). In this case, the H addition during growth not only supports the Mg solubility, but is also essential to suppress the intrinsic O vacancy compensation mechanism. Even though neither scenario creates significant *p*-type conductivity, the doping type conversion and associated drop of the Fermi level present important opportunities for device design, allowing to create a significant built-in field in a *p-n* junction with an adjoining *n*-type material, and greatly suppressing the free electron density. The specific predictions of suitable synthesis process conditions presented in this paper could guide the fabrication of current blocking layers, for example in normally-off (enhancement-mode) vertical $Ga_2O_3$ based MOSFET and as a guard ring for edge termination in such transistors, and in Schottky barrier diodes with increased breakdown voltage.

*Computational details*: The first-principles calculations were performed using the projector augmented wave (PAW) method[70] as implemented in the VASP (Vienna Ab-initio Simulation Package) code[71] for DFT[72], hybrid-DFT[73], and *GW*[74] calculations. The generalized gradient approximation (GGA) of Ref[75] was used for DFT exchange and correlation and the HSE06[76,77] functional for hybrid functional calculations. The defect formation energies and charge transition levels (in both DFT and hybrid-DFT) are calculated in 160 atom defect supercells with a Γ centered 2×2×2 k-mesh, using our recently developed automated defect framework[78]. The PAW potentials "Ga_d", "O", "O_s", "Mg_pv", and "H" of the VASP 4.6 distribution were used. For structures without hydrogen, the soft "O_s" potential was employed with a plane wave energy cutoff of 340 eV. Previous tests have confirmed that such calculations are accurate for sufficiently large interatomic distances.[49] Due to the presence of short O-H bonds for which the soft potential may not be accurate, we used the standard "O" potential with 520 eV cutoff for structures with hydrogen. Finite size corrections are applied as described in Ref.[79] In the hybrid-DFT calculations, the HSE06 funcational[76,77] is employed with α = 0.3 and μ = 0.2 Å$^{-1}$ for the fractional Fock exchange and the range separation parameters, respectively. For the DFT bandgap correction, *GW* calculations[80] gives the individual band edge shifts of $\Delta E_{VBM}$ = −1.82 eV and $\Delta E_{CBM}$ = +1.1 eV relative to DFT-GGA, resulting in the bandgap value of 4.96 eV. The FERE reference elemental chemical potentials of Ref[50] are used for Ga, O, and Mg. For hydrogen, we determined here the value $\mu_H^0$ = −3.49 eV by fitting against the tabulated formation enthalpies[53] of the following compounds, $H_2O$ (ice I$_h$), LiH, $MgH_2$, LiOH and $Mg(OH)_2$. The temperature dependence of the CBM was taken from Ref[23], where it was determined from molecular dynamics simulations.

**Acknowledgements:** This work was funded by the U.S. Department of Energy (DOE), through the Laboratory Directed Research and Development Program of the National Renewable Energy Laboratory (NREL), and by the DOE Advanced Manufacturing Office of the Office of Energy Efficiency and Renewable Energy (EERE). The Alliance for Sustainable Energy, LLC, operates and manages NREL under contract DE-AC36-08GO28308. This work used High-Performance Computing resources at NREL, sponsored by DOE-EERE. The views expressed in the article do not necessarily represent the views of the DOE or the U.S. government.